\documentstyle[preprint,aps,epsf,graphicx]{revtex}
\begin{document}

\tightenlines

\title{Finite temperature effects on the $\bar{K}$ optical potential}

\author{Laura Tol\'os, A.Polls, A.Ramos}
\address{Departament d'Estructura i Constituents de la Mat\`eria,
Universitat de Barcelona, \\
Diagonal 647, 08028 Barcelona, Spain
}

\date{\today}

\maketitle
\begin{abstract}
By solving the Bethe-Goldstone equation, we have obtained
the
$\bar{K}$ optical potential from the $\bar{K}N$ effective interaction in
nuclear matter at $T=0$. We have extended the model by incorporating
finite
temperature effects in order to adapt our calculations to the experimental
conditions in heavy-ion collisions. In the rank of densities ($0-2\rho_0$),
 the finite temperature $\bar{K}$ optical potential shows a smooth
behaviour if we compare it to the $T=0$ outcome.
Our model has also been applied to the study of the ratio between $K^+$
and $K^-$ produced at GSI with $T$ around $70$ MeV. Our results
point at the necessity of introducing an attractive  $\bar{K}$ optical
potential.            

\end{abstract}

\vspace{0.5cm}


\section{The model at $T=0$}

The $\bar{K}$ optical potential in nuclear matter is computed using a
$\bar{K}N$ effective interaction, G-matrix, derived microscopically
from a
meson-exchange potential \cite{Muller90}. The G-matrix is  given 
by the solution of the Bethe-Goldstone equation

\begin{equation}
G(w)=V + V \frac{Q}{w-H_0+i\eta} G(w),
\end{equation}
which takes into account the different channels  ($\bar{K}N$,
$\pi\Sigma$, $\pi\Lambda$), which are coupled by the strong
interaction.

At lowest order in the Brueckner-Hartree-Fock (BHF) theory, the
single-particle potential for
antikaons $U_{\bar{K}}$ at $T=0$ is given by

\begin{eqnarray}
U_{\bar K}(k,E_{\bar K}^{qp})= \sum_{N \leq F} \langle
\bar
K N \mid
 G_{\bar K N\rightarrow
\bar K N} (w = E^{qp}_N+E^{qp}_{\bar K}) \mid \bar K N \rangle,
\end{eqnarray} 
in which  $E^{qp}_{\bar{K}}$ is  self-consistently determined.

In order to introduce in-medium effects on the intermediate states of
the G-matrix, we have employed mean field single-particle propagators for 
the baryons. The
single-particle potential for nucleons has been derived from a 
$\sigma-\omega$ model, while for $\Lambda$ and $\Sigma$, we have
followed the parameterization of Ref. \cite{Balberg97}. Finally, the
pion is dressed with the momentum and energy dependent self-energy of
Ref. \cite{Oset00}, that incorporates the coupling to particle-hole,
$\Delta$-hole and two particle-two hole excitations.

Fig.1  shows the real and imaginary part of $U_{\bar{K}}$ as
a function of
$k_{\bar{K}}$ at
$T=0$ for $\rho=\rho_0$ including or not the dressing of pions.
The long-dashed line has been calculated determining the complex
$\bar{K}$ single particle energy self-consistently. However,
the pion
self-energy in the intermediate states is not included. This procedure is 
similar to the one presented
in Ref. \cite{Tolos01} with the inclusion here of a $\sigma-\omega$ 
model for the nucleon dressing. The presence of the $\Lambda(1405)$
resonance,  dynamically
generated in our model, introduces a strong energy dependence on the
$\bar{K}N$ amplitude which governs, in turn, the behaviour of
$U_{\bar{K}}$. The Pauli blocking on the intermediate nucleons
has a repulsive effect on the resonance, produced by having moved the threshold
of intermediate states to higher energies.
The self-consistent incorporation of the $\bar{K}$ properties on the
$\bar{K}N$ G-matrix moves the resonance back in energy, but it dilutes as  
the density increases.
If we now include the dressing of pions (solid line), we obtain a less 
attractive real
part, going from $-84$ MeV to $-73$ MeV at $k=0$ MeV, and
a smoother
behaviour for the imaginary part. This can be understood by looking again
to  the  $\bar{K}N$ amplitude, specially for $L=0$. Due to the additional
dressing of the pions,
the resonance dissolves even faster loosing structure, although it
shows a
tendency to move to higher energies.

\section{Temperature effects}

The introduction of temperature in the calculation affects the Pauli
blocking of the intermediate nucleon states in the G-matrix. In addition,
the $\bar{K}$ optical potential is calculated according to

\begin{eqnarray}
U_{\bar K}(k,E_{\bar K}^{qp})=
\int  d^3k \ n(k,T) \
\langle \bar K N \mid
 G_{\bar K N\rightarrow \bar K N} (w,T) \mid \bar K N \rangle
\label{uk}
\end{eqnarray}
where $n(k,T)$ is the nucleon momentum distribution at finite temperature.
By looking at Fig.2, one observes that, as
the temperature increases, $U_{\bar K}$ is
less attractive and shows a smoother behaviour in momentum. At finite
$T$, the sum over momenta in Eq.(\ref{uk}) is
extended over the corresponding Fermi distribution, and higher nucleon
momentum translates
into a  weaker interaction. Moreover, the effective interaction has
also changed. However, this last effect is less important. 
At
$T=70$ MeV, the momentum dependence is very smooth, reducing in a factor of
three the difference between low and high momentum. However,  $U_{\bar K}$
is still attractive. This fact can be used to understand the enhancement
of the observed ratio $K^-/K^+$ in heavy-ion collisions at GSI energies.

\section{Ratio  $K^-/K^+$}

{\it Heavy-ion collisions} provide a unique possibility to create a
dense
and hot nuclear system to study the in-medium  properties of hadrons, like
in-medium effects on the $\bar{K}$. The measured ratio of particles 
$K^-/K^+$ gives us
information about the necessity of an in-medium attractive potential for
the
$K^-$, already corroborated in the analysis of kaonic atoms.

In direct nucleon-nucleon collisions, the production of $K^+$ and $K^-$ is
governed by quite distinct thresholds
(for $NN \rightarrow K^+ \Lambda N$, the
threshold
is at 1.58 GeV and for $NN \rightarrow K^+K^-NN$ is 2.5 GeV), and the 
$K^+$ multiplicity
exceeds the $K^-$ one by 1-2 orders at a given energy above thresholds.
However,
this large difference disappears for nucleus-nucleus collisions
($C+C$, $Ni+Ni$ \cite{Barth97-Laue99,Menzel00}), where the data nearly
fall
on
the same curve. Furthermore, we observe that this ratio is approximately
constant for $C+C$, $Ni+Ni$ and $Au+Au$
\cite{Barth97-Laue99,Menzel00,Forster}, although absorption of $K^-$ via
$K^- N \rightarrow Y \pi$ is supposed to be higher for heavier nuclei
\cite{Senger01}.
It turns out that both effects can be interpreted by assuming an in-medium
attractive $U_{\bar{K}}$ \cite{Cassing97-Brat97,Li97}


Analysis of  $K^+/K^-$ in $Ni+Ni$ at GSI energies
\cite{Barth97-Laue99,Menzel00}
gives a ratio around $30$. In Ref. \cite{Cleymans}, it was shown
 that this ratio  can be explained in terms of a {\it
thermal model}, i.e., the final state can be described as 
a {\it hadronic gas} in chemical equilibrium at a given
temperature. Imposing exact strangeness
conservation, one obtains,

\vspace{0.2cm}
\begin{eqnarray}
\frac{K^+}{K^-} =
\frac{g_{K^+} V \int \frac{d^{3}p}{(2 \pi)^3} e^{\frac{-E_{K^+}}{T}}
(g_{K^-} V \int \frac{d^{3}p}{(2 \pi)^3} e^{\frac{-E_{K^-}}{T}}+
g_{\Lambda} V \int \frac{d^{3}p}{(2 \pi)^3}
e^{\frac{-E_{\Lambda}}{T}+\frac{\mu_B}{T} }+
g_{\Sigma} V \int \frac{d^{3}p}{(2 \pi)^3}
e^{\frac{-E_{\Sigma}}{T}+\frac{\mu_B}{T} }
)}
{g_{K^-} V \int \frac{d^{3}p}{(2 \pi)^3} e^{\frac{-E_{K^-}}{T}}
(g_{K^+} V \int \frac{d^{3}p}{(2 \pi)^3} e^{\frac{-E_{K^+}}{T}})
},
\end{eqnarray}
where $g_{\alpha}$ are the spin-isospin degeneracies and $V$ is the
interaction volume. The above expression shows that the presence of $K^+$
must
be compensated by $\Lambda$, $\Sigma$ and
$K^-$ to conserve strangeness equal to zero, while $K^-$ can only
be compensated by $K^+$.

It is found that the data \cite {Cleymans98} can be explained using a
$T=70 \pm 10$ MeV and $\mu_B= 720 \pm 20$ MeV. G.Brown \cite{Brown01}
introduced the notion of "broad-band equilibration" in heavy-ion processes
at GSI energies. In this interpretation, due to the compensation between
the increase in $\mu_B$ as the density increases and the density
dependence of $U_{\bar{K}}$, the ratio is constant over a large range
of densities, not only for $\rho \sim \frac{1}{4} \rho_0$, but also up to
$2 \rho_0$.

In Fig.3, we show the importance of dressing $K^-$.
Plotting the
ratio $K^+/K^-$ as function of density, it can be seen that the effect
of
including $U_{\bar{K}}$ reduces the ratio that, otherwise, would
increase steadily (short-dashed line). It is also seen that dressing the
pions
induces
appreciable differences in the  calculations (long-dashed line).

Once  the full dressing is included, Fig.4 shows
the sensitivity of the ratio  to the chemical potential, and hence to
density, as well as the importance of introducing the $\Sigma$ in the
balance equation. The
"broad-band equilibration" is hardly reproduced  when
the $\Sigma$ is taken into account, and one obtains large differences
depending on the character (attractive or repulsive) of the $\Sigma$
optical potential.

\section{Conclusions}

Performing a microscopic calculation of the Brueckner-Hartree-Fock type
, we conclude that dressing pions in the
calculation of $U_{\bar{K}}$ introduces significant differences, around 13$\%$ at zero momentum. Moreover, the introduction of temperature gives
rise
to a smoother behaviour in momentum, being  attractive even at
momentum as large as
500 MeV, relevant  in heavy-ion collisions. Heavy-ion
collisions are the perfect scenario to test in-medium $\bar{K}$
properties, specially in understanding the ratio $K^-/K^+$. The
attempts to reproduce this ratio require an attractive
$U_{\bar{K}}$. The ratio shows a  measurable density dependence.
 The
"broad-band equilibration" is far from being reproduced if 
hyperons like  $\Sigma$ are taken into account, showing large differences
with regard to only including $\Lambda$ hyperons.

\section*{Acknowledgments}

We are very grateful to Dr. J\"urgen Schaffner-Bielich for the useful
discussions that have made possible this work and L.T. wishes to 
acknowledge Brookhaven National Laboratory for kind hospitality during her
stay.

\begin{figure}[htb]
\centerline{
     \includegraphics[width=0.45\textwidth]{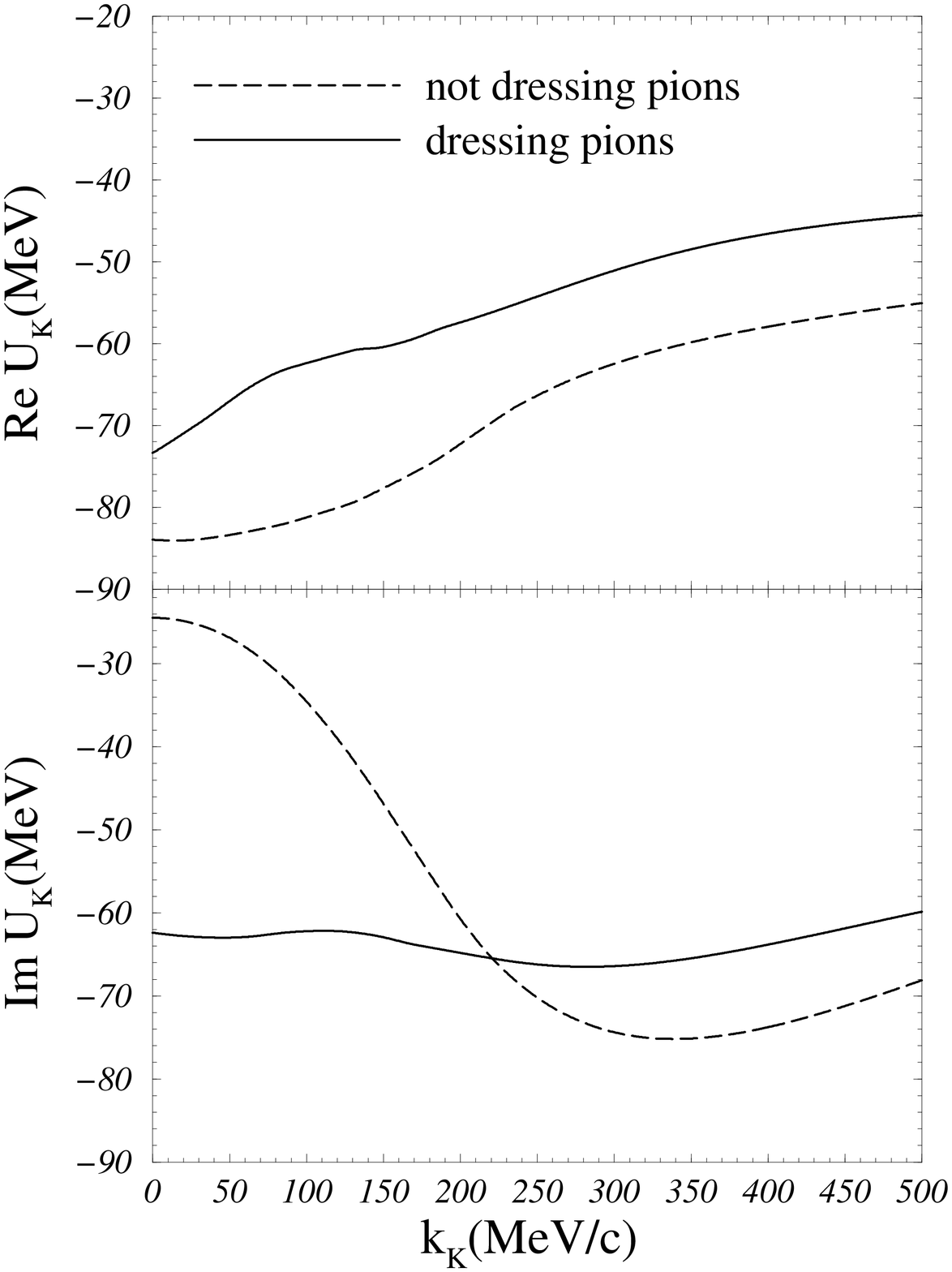}
}
      \caption{\small
$U_{\bar{K}}$ as a function of $k_{\bar{K}}$ for $T=0$,
including or not the dressing of pions at $\rho=0.17 fm^{-3}$
}
        \label{fig:fig1}
\end{figure}

\begin{figure}[htb]
\centerline{
     \includegraphics[width=0.45\textwidth]{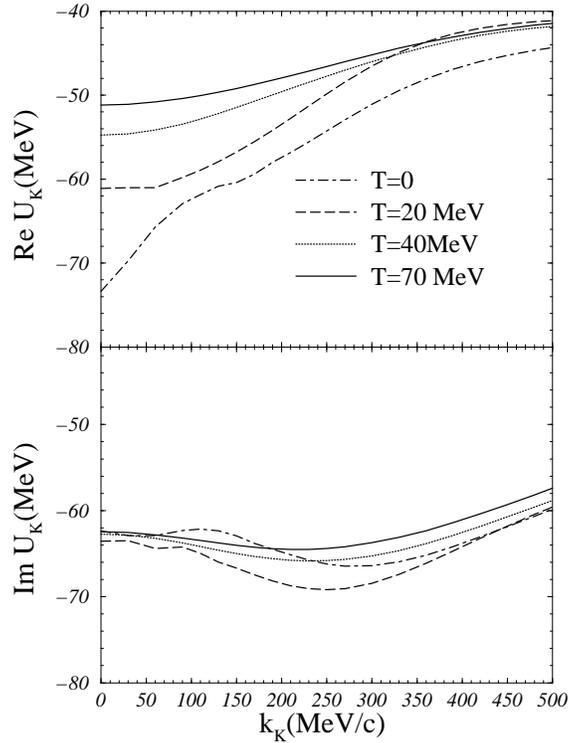}
}
      \caption{\small
$U_{\bar{K}}$ as a function of $k_{\bar{K}}$ for different $T$,
at $\rho=0.17 fm^{-3}$
}
        \label{fig:fig2} 
\end{figure}

\begin{figure}[htb]
\centerline{
     \includegraphics[width=0.6\textwidth]{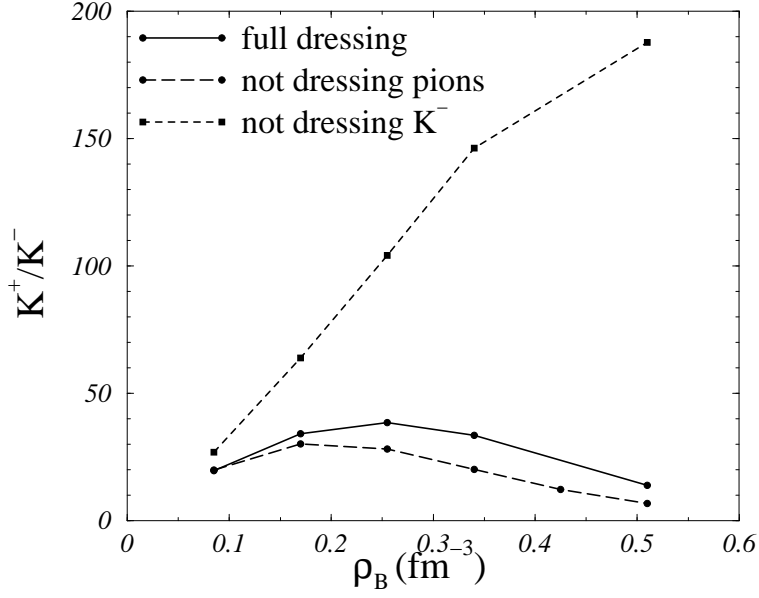}
}
      \caption{\small
$K^+/K^-$ as a function of density at $T=70MeV$, including
or not $U_{\bar{K}}$
}
        \label{fig:fig3}
\end{figure}

\begin{figure}[htb]
\centerline{
     \includegraphics[width=0.6\textwidth]{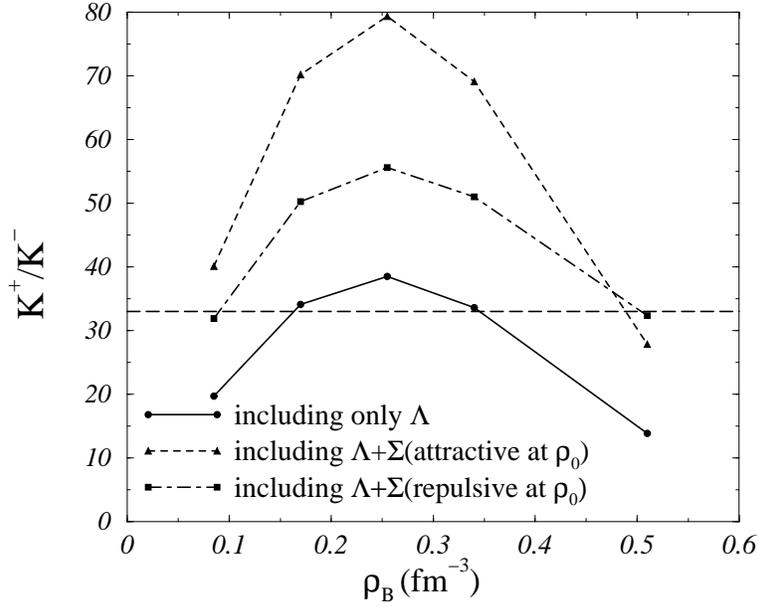}
}
      \caption{\small
$K^+/K^-$ as a function of density at $T=70MeV$, included
$U_{\bar{K}}$, including or not $\Sigma$
}
        \label{fig:fig4}
\end{figure}

\end{document}